\documentclass[letterpaper, 10 pt, conference]{ieeeconf}  
\usepackage{amsmath,amsfonts}
\usepackage{algorithmic}
\usepackage{algorithm}
\usepackage{array}
\usepackage[caption=false,font=normalsize,labelfont=sf,textfont=sf]{subfig}
\usepackage{textcomp}
\usepackage{stfloats}
\usepackage{url}
\usepackage{verbatim}
\usepackage{graphicx}
\usepackage{cite}
\usepackage{units}
\usepackage{dsfont}
\usepackage{comment}
\usepackage{booktabs}
\usepackage{marginnote}
\usepackage[dvipsnames]{xcolor}
\usepackage[official]{eurosym}
\usepackage{xfrac}
\usepackage{array}

\usepackage{amsthm}

\newtheorem{theorem}{Theorem}[section]

\newcounter{thm}
\newtheorem{prob}[thm]{Problem}

\IEEEoverridecommandlockouts                              

\newif\ifmargincomments 
\margincommentstrue

\ifmargincomments

\else

\fi

\newcommand{\revVPPC}[1]{{\leavevmode\color{black}#1}}

\title{\LARGE \bf Geometric Scaling Laws for Axial Flux Permanent Magnet Motors in In-Wheel Powertrain Topologies}

\author{Olaf Borsboom$^{1}$, Arnab Bhadra$^{1}$, Mauro Salazar$^{1}$ and Theo Hofman$^{1}$
\thanks{This publication is part of the project NEON with project number 17628 of the research program Crossover, which is (partly) financed by the Dutch Research Council (NWO).}
\thanks{$^{1}$Department of Mechanical Engineering, Control Systems Technology,
        Eindhoven University of Technology, 5600 MB  Eindhoven, The Netherlands. 
        E-mail: {\tt\small o.j.t.borsboom@tue.nl}}%
}

\begin{document}

\maketitle
\thispagestyle{empty}
\pagestyle{empty}

\begin{abstract}
In this paper, we present geometric scaling models for axial flux motors (AFMs) to be used for in-wheel powertrain design optimization purposes.
\revVPPC{We first present a vehicle and powertrain model, with emphasis on the electric motor model.
We construct the latter by formulating the analytical scaling laws for AFMs, based on the scaling concept of RFMs from the literature, specifically deriving the model of the main loss component in electric motors: the copper losses. 
We further present separate scaling models of motor parameters, losses and thermal models, as well as the torque limits and cost, as a function of the design variables.
Second, we validate these scaling laws with several experiments leveraging high-fidelity finite-element simulations.
Finally, we define an optimization problem that minimizes the energy consumption over a drive cycle, optimizing the motor size and transmission ratio for a wide range of electric vehicle powertrain topologies. }
In our study, we observe that the all-wheel drive topology equipped with in-wheel AFMs is the most efficient, but also generates the highest material cost.
\end{abstract}

\section{Introduction}\label{sec:introduction}

To advance the adoption of electric vehicles, it is important to holistically design the powertrain.
This is a challenging task, since the powertrain is a highly interconnected and multidisciplinary system.
One of the most demanding components is the electric motor (EM), especially when engineers are faced with the task of translating vehicle requirements to a well-performing motor design.
If the motor is oversized, it will take up valuable space in the chassis and be unnecessarily heavy. 
Yet if we undersize the motor, the vehicle might not be able to meet the performance requirements (in terms of top speed or acceleration) it has been set for, and be prone to overheating.

Zooming in on the EM, one prospective technology is the axial flux permanent magnet synchronous motor (AFM)~\cite{IndustryARC2021}.
The concept of AFMs has been long known~\cite{Traxial2021} and is promising due to its higher torque density and efficiency compared to their counterparts radial flux motors (RFMs)~\cite{ElFahem2020}, but their application has remained limited due to the relatively high construction complexity, driving up the purchasing cost of the vehicle.

\revVPPC{To optimize the design of EMs of any type, we need EM models.
However, there exists an inherent trade-off in modeling between accuracy, level of detail, and the number of design variables on the one hand, and computational speed and powertrain/system-level compatibility on the other hand.}
This calls for methods to optimally size AFMs with models that are accurate but tractable for holistic powertrain design.

\subsubsection*{Related literature}
Our investigation primarily pertains to two main research streams. 
The first stream deals with holistic powertrain design, with emphasis on the scalability of EM models.
Geometrically scalable EM models have been introduced and validated in~\cite{StipeticZarkoEtAl2015} and~\cite{StipeticZarkoEtAl2016}, and employed for powertrain design optimization in~\cite{RamakrishnanStipeticEtAl2018,BorsboomSalazarEtAl2022, BorsboomLokkerEtAl2023}. 
These methods have proven to be useful, accurate and computationally efficient, but have not been applied to AFMs yet. 

The second research stream focuses on analytical modeling for AFMs, discussed in~\cite{HincapieSandovalEtAl, SahinVandenputEtAl2001, Zhilichev1998, KurronenPyrhoenen2007}. 
This holds for both 2D and (quasi-)3D models and different combinations of inputs and outputs, but the influence of sizing or geometrically scaling is not captured in these models. 

To conclude, to the best of the authors’ knowledge, there are no studies performed to date, exploring the scaling laws of AFMs that are tractable to design optimization.

\begin{figure}
	\centering
	\includegraphics[width=\columnwidth]{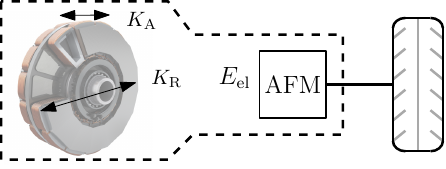}
	\caption{The axial flux motor (AFM) in-wheel powertrain topology for one wheel. This simple powertrain is to be duplicated to two wheels in front and rear-wheel drives, and to four wheels in an all-wheel drive. The AFM design is proportionally scaled in axial and radial direction with scaling factors $K_\mathrm{A}$ and $K_\mathrm{R}$, respectively. AFM picture taken from~\cite{GCC2021}.}
	\label{fig:scaling}
\end{figure}%

\subsubsection*{Statement of Contributions}
In this paper, we introduce analytical scaling laws that capture the behavior of geometrically scaling a referent AFM design. 
Specifically, we \revVPPC{derive models for the phase current, phase resistance and copper losses, and concisely present models for the torque limits, inductances, motor losses, and heating as a function of the scaling factors.}
\revVPPC{We validate the derived scaling equations with high-fidelity, finite-element (FE) simulation models of synchronous motors on multiple levels with different experiments, namely on the parameter and loss level for a fixed operating point, on the loss level under drive cycle operation and on the efficiency level for multiple operating points.}
We also present a design optimization study for a small range of electric powertrain topologies.

\subsubsection*{Organization}
This paper is organized as follows:
\revVPPC{Section~\ref{sec:methodology} presents vehicle and powertrain model, the scaling laws and models for motor parameters and the thermal network, and the summarizing design optimization problem.}
Section~\ref{sec:results} displays the results by showcasing the presented framework on a drive cycle and a finite set of powertrain topologies.
The conclusions are drawn in Section~\ref{sec:conclusions}.

\section{Methodology}\label{sec:methodology}
This section introduces the vehicle and powertrain models, with \revVPPC{emphasis} on the motor scaling procedure and its application on motor parameters.
Subsequently, we present the validation, followed by the summarizing optimization problem and a discussion of the proposed method.

\subsection{Longitudinal Vehicle Dynamics}\label{subsec:longvehdyn}
We adopt the quasi-steady state vehicle modeling approach, which is well known in powertrain design.
We calculate the required wheel torque along the drive cycle \revVPPC{$T_\mathrm{req}(t)$} on a flat road by
\begin{multline}
	T_\mathrm{req}(t) = r_\mathrm{w} \cdot \left( m_\mathrm{v} \cdot a(t) + \vphantom{\frac{1}{2}} \right. \\
	 \left.  \frac{1}{2}\cdot \rho_\mathrm{a} \cdot c_\mathrm{d} \cdot A_\mathrm{f} \cdot v(t)^2 + c_\mathrm{r} \cdot m_\mathrm{v} \cdot g \right),
\end{multline}
where \revVPPC{$v(t)$ and $a(t)$} are the velocity and acceleration of the vehicle during the drive cycle, respectively.
Hereby, $r_\mathrm{w}$ is the radius of the wheels, $m_\mathrm{v}$ is the mass of the vehicle, $\rho_\mathrm{a}$ is the density of air, $c_\mathrm{d}$ is the aerodynamic drag coefficient, $A_\mathrm{f}$ is the frontal area of the vehicle, $c_\mathrm{r}$ is the rolling resistance coefficient, and $g$ is the gravitational constant. 
We omit time-dependence whenever it is clear from the context.

\subsection{Transmission}\label{subsec:gb}
\revVPPC{The one-speed transmission is modeled with a fixed efficiency $\eta_\mathrm{g}$ and gear ratio $\gamma$.} 
The motor torque $T_\mathrm{m}$ is equal to
\begin{align}
	T_\mathrm{m} =
	\begin{cases}
		\frac{r_\mathrm{b} \cdot T_\mathrm{req}}{\gamma \cdot \eta_\mathrm{g}} \quad &\text{ if } T_\mathrm{req} \geq 0 \\
		\frac{r_\mathrm{b} \cdot T_\mathrm{req}\cdot \eta_\mathrm{g}}{\gamma} \quad &\text{ if } T_\mathrm{req} < 0,
	\end{cases} 
\end{align}
where $r_\mathrm{b}$ is the regenerative braking fraction, depending on the powertrain layout (front, rear, or all-wheel drive).
The EM speed $\omega_\mathrm{m}$ is given by
\begin{equation}
	\omega_\mathrm{m} = \gamma \cdot \frac{v}{r_\mathrm{w}}. 
\end{equation}

\subsection{Electric Motor}\label{subsec:em}
The \revVPPC{visualization} of scaling is shown in Fig.~\ref{fig:scaling}. 
We scale the referent AFM in two directions: axial direction and radial direction. 
With axial scaling, we stretch the core length of the motor with factor $K_\mathrm{A}$, while keeping the lamination cross-section, and the relative lengths of the stator rotor and magnet lengths unchanged. 
When we scale radially, we proportionally change all component dimensions with factor $K_\mathrm{R}$, while keeping the slot current density unchanged. 
This results in scaling laws for all relevant motor characteristics:
the volume $V$\revVPPC{, viewing the motor as a cylinder};
the mass, which is calculated through the material density;
the cost, which is calculated through the specific cost (in \unitfrac{\euro{}}{kg}) for each material;
the nominal torque;
the copper, iron, bearing and windage losses, $P_\mathrm{Cu}$, $P_\mathrm{Fe}$, $P_\mathrm{br}$, and $P_\mathrm{wind}$, respectively, modeled with analytical equations based on motor parameters; and
the heat capacitance and transfer surface areas.
The most important scaling laws are summarized in Table~\ref{tab:scalinglaws}.

\begin{table}[]
	\caption{Scaling Laws for Axial Flux Motors}
	\centering
		\revVPPC{
	\renewcommand{\arraystretch}{1.3}
	\begin{tabular}{@{}lll@{}}
		\toprule
		\textbf{Parameter} & \textbf{Scaling law} & \textbf{Unit} \\
		\midrule
		Torque & $T_\mathrm{m} = K_\mathrm{R}^3 \, K_\mathrm{A} \,  T_\mathrm{m,0}$ & Nm \\
		Power & $P_\mathrm{m} = K_\mathrm{R}^3 \, K_\mathrm{A} \,  P_\mathrm{m,0}$ & W \\
		Phase inductance & $L_\mathrm{dq} =  K_\mathrm{R}^2 \, K_\mathrm{A} \, L_\mathrm{dq,0}$ & H \\
		Magnetic flux & $\psi = K_\mathrm{R}^2 \,K_\mathrm{A} \, \psi_0 $ & Wb \\
		Phase resistance & $ R_\mathrm{ph} = K_\mathrm{R} \, K_\mathrm{A} \, R_\mathrm{ph,0} $ & $\Omega$ \\
		Phase current &  $ I_\mathrm{ph} = K_\mathrm{R} \, I_\mathrm{ph,0} $ & A \\
		Component volume & $V = K_\mathrm{R}^2 \, K_\mathrm{A} \,  V_{0}$  & m$^3$\\
		Copper losses & $P_\mathrm{Cu} = K_\mathrm{R}^3 \, K_\mathrm{A} \,  P_\mathrm{Cu,0}$ & W \\
		\bottomrule
	\end{tabular}
	\label{tab:scalinglaws}
	}	
\end{table}

\begin{figure}
	\centering
	\includegraphics[width=\columnwidth]{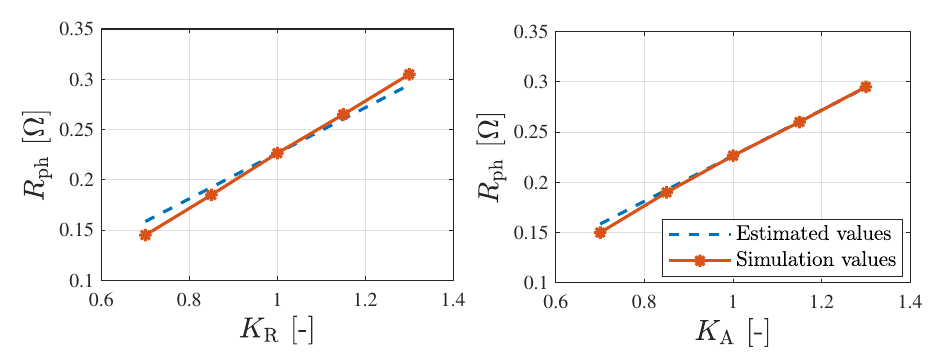}
	\caption{Scaling the phase resistance as a function of $K_\mathrm{R}$ and $K_\mathrm{A}$.}
	\label{fig:Rph}
\end{figure}

\begin{figure}
	\centering
	\includegraphics[width=\columnwidth]{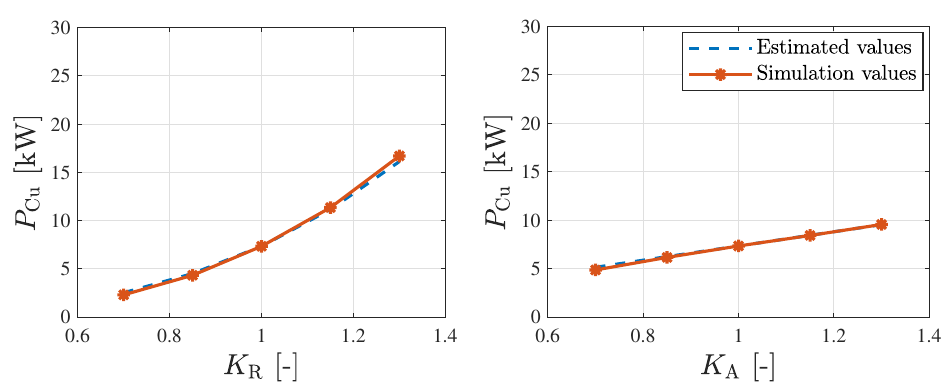}
	\caption{Scaling the copper losses as a function of $K_\mathrm{R}$ and $K_\mathrm{A}$.}
	\label{fig:PCu}
\end{figure}

As an example, we will demonstrate the scaling of the losses occurring in the copper windings. 
The copper losses $P_\mathrm{Cu}$ for a three-phase synchronous motor are typically modeled with
\begin{equation}
	P_\mathrm{Cu} = 3 \cdot I_\mathrm{ph}^2 \cdot R_\mathrm{ph},
\end{equation}
where $I_\mathrm{ph}$ is the phase current and $R_\mathrm{ph}$ is the phase resistance. 
The phase resistance is equal to 
\begin{equation}
	R_{\mathrm{ph}} = \frac{1}{a_\mathrm{p}} \cdot \rho \cdot \frac{l_\mathrm{t} \cdot \frac{N_\mathrm{c}} {a_\mathrm{p}} \cdot \frac{Q_\mathrm{s}}{3}}{\frac{1}{2} \cdot \frac{A_{\mathrm{slot}}}{N_\mathrm{c}}\cdot k_{\mathrm{Cu}}},
\end{equation}
where $a_\mathrm{p}$ is the number of parallel paths, $\rho$ is the material resistivity, $l_\mathrm{t}$ is the mean turn length,  $N_\mathrm{c}$ is the number of coils, $Q_\mathrm{s}$ is the number of stator slots, $A_\mathrm{slot}$ is the slot area, and $k_\mathrm{Cu}$ is the slot fill factor. 
We know that the slot area increases proportionally with $K_\mathrm{A}$ and $K_\mathrm{R}$, the mean turn length grows with $K_\mathrm{R}$, the slot fill factor scales with $\sfrac{1}{K_\mathrm{R}}$ , and the number of coils must multiply with $K_\mathrm{A}$ to maintain the current density. 
This results in the following scaling law:
\begin{equation}
	R_{\mathrm{ph}} = K_\mathrm{R} \cdot K_\mathrm{A} \cdot R_{\mathrm{ph,0}},
\end{equation}
where symbols with the subscript 0 represent the unscaled parameter from the referent motor design. 
This way, $R_\mathrm{ph,0}$ is the unscaled, referent phase resistance.
To maintain the current density, the phase current scales as follows:
\begin{equation}
	I_\mathrm{ph} = K_\mathrm{R} \cdot I_\mathrm{ph,0}.
\end{equation}
This results in the scaling law for the copper losses:
\begin{equation}
	P_{\mathrm{Cu}} = 3 \cdot \left(I_{\mathrm{ph,0}} \cdot K_\mathrm{R}^\mathrm{2} \right) \cdot \left(R_{\mathrm{ph,0}} \cdot K_\mathrm{R}\cdot K_\mathrm{A}\right).
\end{equation}
\revVPPC{The implementation of} scaling laws on a referent AFM design for the phase resistance and copper losses are given in Fig.~\ref{fig:Rph} and Fig.~\ref{fig:PCu}, respectively.
Following similar reasoning for the other losses, of which the full derivations will be given in the extended version of this paper, the total electric input power to the motor $P_\mathrm{el}$ is equal to
\begin{equation}
	P_\mathrm{el} = T_\mathrm{m} \cdot \omega_\mathrm{m} + P_\mathrm{Cu} + P_\mathrm{Fe} + P_\mathrm{br} + P_\mathrm{wind},
\end{equation}
resulting dynamics of the electric energy provided to the EM, $E_\mathrm{el}$:
\begin{equation}
	\frac{\mathrm{d}}{\mathrm{d}t}E_\mathrm{el} = P_\mathrm{el}.
\end{equation}

\subsubsection*{Thermal Model}
\begin{figure}
	\centering
	\includegraphics[width=0.8\columnwidth]{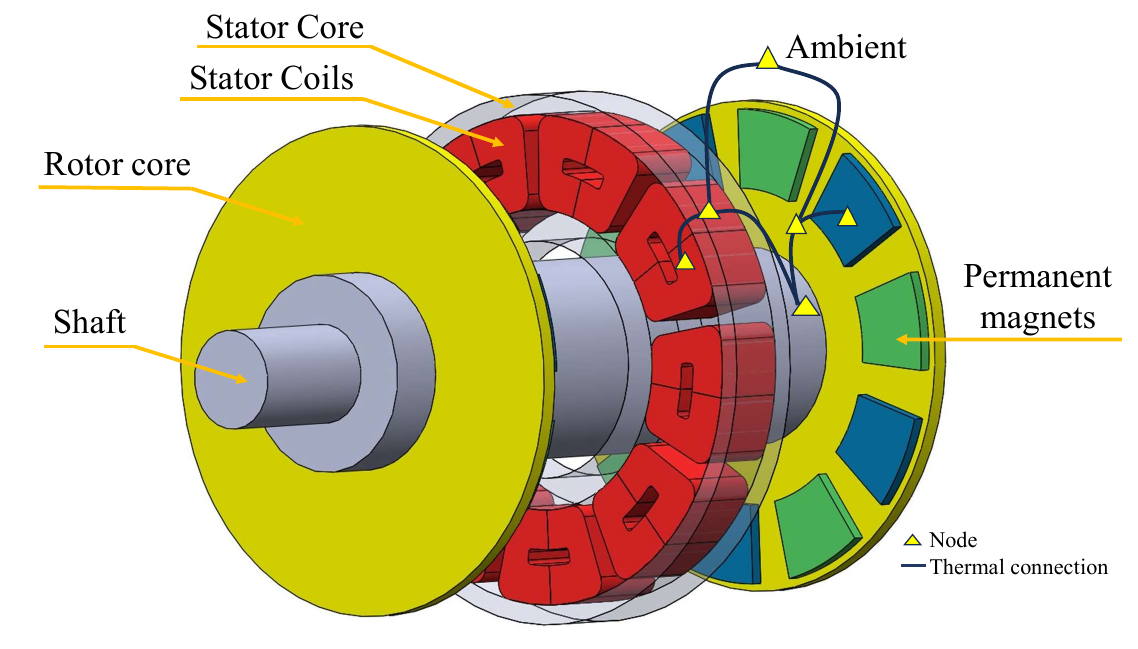}
	\caption{The assembly of the AFM with the thermal network, given in nodes and thermal connections.}
	\label{fig:AFMassembly}
\end{figure}

\begin{figure}
	\centering
	\includegraphics[width=\columnwidth]{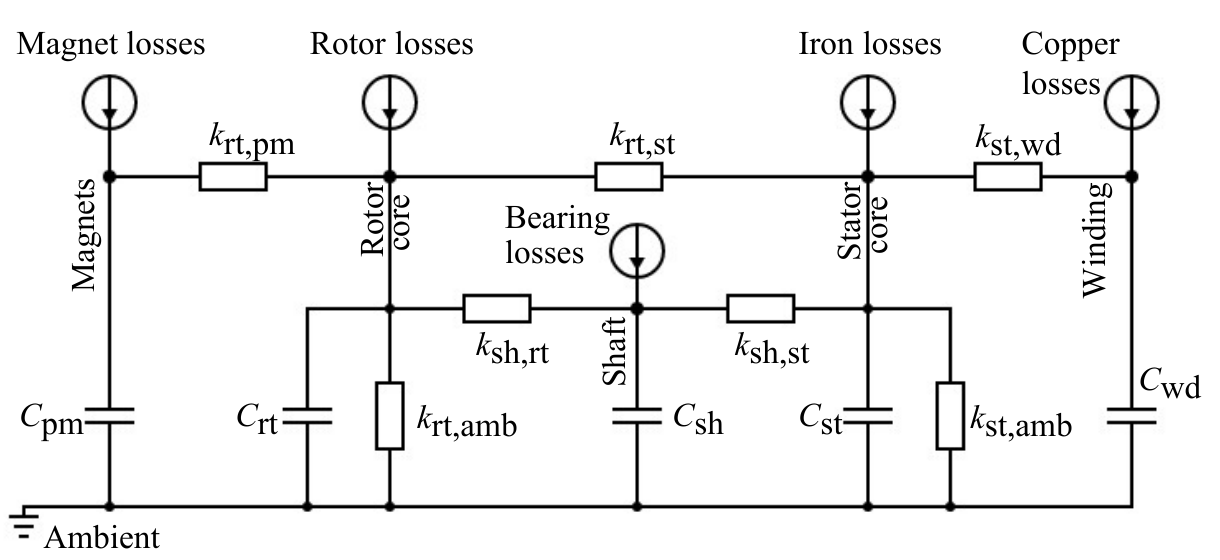}
	\caption{The LPTN of the referent AFM design.}
	\label{fig:LPTN}
\end{figure}

The thermal behavior of the motor is estimated with a Lumped-Parameter Thermal Network (LPTN). 
The physical layout of the AFM is given in Fig.~\ref{fig:AFMassembly}, while the derived LPTN is shown Fig.~\ref{fig:LPTN}.
From the LPTN, we can write a system of heat flow equations with each of the nodes representing a component. 
The scaling factors impact the component volumes, the contact areas between the components, and the linear velocity at the outer radius of the rotor. 
These, in turn, influence the heat capacities, heat transfer coefficients, and convective behavior.
With this information, we can update the LPTN heat flow system.
\revVPPC{The full derivation will be given in the extended version of this paper.}

\subsection{Validation}\label{subsec:val}
 The validation process takes place in the high-fidelity \revVPPC{FE} computer-aided design tool SolidWorks, together with the Electro-Magnetic-Simulation package\revVPPC{, and on multiple levels of the EM models.}
The computer-aided design model of an AFM is shown in Fig.~\ref{fig:AFMassembly}.

\revVPPC{As a validation on the parameter level and the loss level, the scaling law predictions and FE simulation values of the phase resistance and copper loss for a fixed operating point are given in Fig.~\ref{fig:Rph} and Fig.~\ref{fig:PCu}, respectively.}
All validation simulations are run at nominal torque at \unit[1000]{rpm}. 
The measured scaling factors are $K_\mathrm{R}, K_\mathrm{A} \in \left\{0.7,\ 0.85,\ 1.15,\ 1.3\right\}$. 

\revVPPC{Next, we take a scaled design of both EM technologies, namely $K_\mathrm{R} = 1.13$, $K_\mathrm{A} = 1.13$ for the AFM, and $K_\mathrm{R} = 0.71$, $K_\mathrm{A} = 1.21$ for the RFM, and perform two additional validation analyses.
First, we simulate the motor and the powertrain on a drive cycle, and compare the copper and iron losses predicted with the scaling laws and simulated with the FE tool.
We use the normalized root-mean-squared error (NRMSE) metric for the comparison.
Second, we select three operating points (OP) of the EM (OP1: low speed, high torque; OP2: medium speed, medium torque; OP3: high speed, low torque), and compare the efficiency calculated with the scaling laws and the FE tool in percentage-points.
The numerical results are given in Table~\ref{tab:validation}.
What can be observed is that all loss power errors are below 8\%, and in particular the iron losses can be predicted accurately.
The higher errors in the copper losses can be attributed to the effects in the end-windings. 
Nonetheless, the difference in efficiency on the operating points, which is the most interesting metric in the context of powertrain design, is well below 1\%pt and often minor.}



\begin{table}[!t]
	\caption{Validation of Scaled EMs \\(AFM: $K_\mathrm{R} = 1.13$, $K_\mathrm{A} = 1.13$, RFM: $K_\mathrm{R} = 0.71$, $K_\mathrm{A} = 1.21$)}
	\centering
	\renewcommand{\arraystretch}{1.3}
	\begin{tabular}{@{}lccc@{}}
		\toprule
		\textbf{EM Technology} & \multicolumn{2}{c}{\textbf{Losses NRMSE (\%)}} & \textbf{Efficiency Diff. (\%pt)} \\
		\cmidrule(lr){2-3} \cmidrule(lr){4-4}
		& \textbf{Copper} & \textbf{Iron} & \textbf{OP1} / \textbf{OP2} / \textbf{OP3} \\
		\midrule
		AFM & 3.48 & 0.62 & 0.73 / 0.55 / 0.15 \\
		RFM & 7.93 & 1.21 & 0.33 / 0.03 / 0.03 \\
		\bottomrule
	\end{tabular}
	\label{tab:validation}
\end{table}

\subsection{Optimization Problem}\label{subsec:optprob}
Given these scaling laws, we can find the optimal EM size by solving the optimization problem that minimizes the EM input energy usage $E_\mathrm{el}(T)$ over a drive cycle at the final time $T$, with the scaling factors $K_\mathrm{A}, K_\mathrm{R}$ and transmission ratio \revVPPC{$\gamma$} as design variables $p$, the power distribution (in case of an all-wheel drive) as control variable $u$, subject to constraints related to thermal effects, performance, and design limitations. 

\begin{prob}[EM Design Problem]\label{prob:main}
	The optimal EM design is the solution of
	\begin{equation*}
		\begin{aligned}
			&\!\min & &E_\mathrm{el}(T) =  \int_{0}^{T} P_\mathrm{el}(t) \mathrm{d}t \\
			& \textnormal{s.t. } & &\textnormal{Design Variable Constraints}\\
			& & &\textnormal{Input Variable Constraints}\\
			& & &\textnormal{Performance Constraints}.
		\end{aligned}
	\end{equation*}
\end{prob}

\subsection{Discussion}\label{subsec:discussion}
A few comments are in order.
First, we only select continuous optimization variables to proportionally scale the referent motor design.
The framework could be expanded by individually rescaling certain design variables, such as the air gap, or by including integer variables, such as the number of pole pairs and slots, penalizing simplicity.
Second, there are no guarantees on global optimality, but we believe that the resulting designs are prospective (for instance as initial guesses for further manual fine-tuning by EM design experts), since the solutions are coherent for multiple topologies (see Section~\ref{sec:results}).
Third, we have not validated the scaling laws on the motor level, by, for instance, comparing the full efficiency map generated by our scaling laws with one produced by high-fidelity tools or provided on a data sheet.
Although this is a high priority for future work, the validations on the loss level and three predetermined points in the efficiency map already are promising.

\section{Results}\label{sec:results}
\begin{figure}[]
	\centering
	\includegraphics[width=\columnwidth]{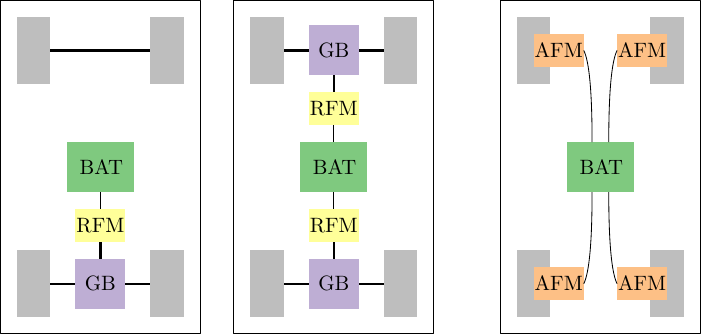}
	\caption{\revVPPC{The tested powertrain topologies. The two left topologies are central drive RFM topologies, namely a rear-wheel and an all-wheel driven one. The right topology is an all-wheel driven one with four in-wheel AFMs. All powertrains are equipped with a battery (BAT) and the RFM ones have a final reduction/differential gear box (GB).}}
	\label{fig:topologies}
\end{figure}

We showcase the scaling laws on a small set of powertrain topologies shown in Fig.~\ref{fig:topologies}. 
\revVPPC{We consider an RFM rear-wheel driven (RWD) powertrain, which is the topology that the Volkswagen ID.3~\cite{EVDatabase2024} is equipped with, an RFM all-wheel drive (AWD), and an AFM AWD.}
We assume that the AFMs are only mounted as in-wheel motors with a direct-drive, which means that we connect them directly to the wheels without a transmission. 
The RFM topologies are equipped \revVPPC{with a one-speed transmission in a final reduction-differential unit.} 
The vehicle parameters are taken from the Volkswagen ID.3~\cite{EVspecs2024}, which are presented in Table~\ref{tab:vehparams}, and the selected driving mission for simulation is the WLTP Class 3 cycle.
We compare all considered topologies \revVPPC{with real-world Volkswagen ID.3 performance and consumption information taken from data bases~\cite{EVDatabase2024, EVspecs2024}, which are held as a baseline.
The parametrization of the performance constraints in Problem~\ref{prob:main} is also based on the baseline vehicle specifications, with a small modification in the acceleration time.}
The selected optimization algorithm is the gradient-free Nelder-Mead simplex method, implemented with the Matlab Optimization toolbox.


\begin{table}[]
	\caption{Vehicle and Simulation Parameters}
	\centering
	\revVPPC{\renewcommand{\arraystretch}{1.3}
	\begin{tabular}{@{}llll@{}}
		\toprule
		\textbf{Parameter} & \textbf{Symbol} & \textbf{Value} & \textbf{Unit} \\
		\midrule
		Wheel radius & $r_\mathrm{w}$ & 0.26 & m \\
		Vehicle mass & $m_\mathrm{w}$ & 1700 & kg \\
		Aerodynamic drag coefficient & $c_\mathrm{d}$ & 0.263 & 1 \\
		Frontal area & $A_\mathrm{f}$ & 2.36 & m$^2$ \\
		Rolling resistance coefficient & $c_\mathrm{r}$ & 0.012 & 1 \\
		\midrule
		Air density & $\rho_\mathrm{air}$ & 1.2041 & kg/m$^3$ \\
		Gravitational constant & $g$ & 9.81 & m/s$^2$ \\
		\midrule
		Top speed & $v_\mathrm{max}$ & 160 & km/h \\
		Accel. time (0-100 km/h) & $t_\mathrm{a}$ & 8 & s \\
		\bottomrule
	\end{tabular}
	\label{tab:vehparams}}
\end{table}

\begin{table}[]
	\caption{Numerical Results}
	\centering
	\revVPPC{\renewcommand{\arraystretch}{1.3}
	\setlength{\tabcolsep}{2pt} 
	\begin{tabular}{@{}lccccl@{}}
		\toprule
		\textbf{Metric} & \textbf{RWD} & \textbf{RWD} & \textbf{AWD} & \textbf{AWD} & \textbf{Unit} \\
		& \textbf{RFM} & \textbf{RFM} & \textbf{RFM} & \textbf{AFM} & \\
		& \textbf{(Baseline)} & & & & \\
		\midrule
		$E_\mathrm{el}(T)$ & 3.49 & 3.42 & 3.00 & 2.88 & kWh \\
		Accel. time (0-100 km/h) & 8.9 & 8 & 8 & 8 & s\\
		Top speed & 160 & 163 & 169 & $>$180 & km/h\\
		Volume & 3.5 & 3.3 & 16-1.8 & 1.7-2.6 & m$^3$ \\
		Mass & N/A & 17.8 & 18.4 & 66.5 & kg \\
		Material cost & N/A & 115 & 121 & 393 & \euro{} \\
		\bottomrule
	\end{tabular}
	\label{tab:results}}
\end{table}

The results are given in Table~\ref{tab:results}, where we make a number of observations:
\revVPPC{
First, we see that the AWD-RFM powertrain is 12\% more energy efficient than the RWD-RFM one. 
This can partially be attributed to the enhanced regeneration capabilities of an AWD vehicle w.r.t a RWD one, especially with recuperation on the front axle, where most of the load is transferred to during braking.
Second, we observe that an AWD equipped with AFMs performs 4\% more energy efficient than one equipped with RFMs, due to the intrinsically higher efficiency of the AFM technology.
However, we also notice that the AFMs are sized such that the acceleration constraint is satisfied with adequate torque, which results in an ample satisfaction of the top speed constraint according to our assumptions and calculations.
This leads to our third observation, namely on the cost of the AFMs in the AWD topology.
Due to the required larger sizing of the motors and the inherently higher cost of AFMs, the material cost estimated with our simplified models is more than triple the expenses of the motors in the AWD-RFM powertrain.
}

\section{Conclusions}\label{sec:conclusions}
In this paper, we leveraged analytical equations of axial flux motors (AFMs) to formulate scaling laws of the motor parameters, limits, losses and the thermal behavior as a function of the design variables.
\revVPPC{We validated these scaling laws in several ways: on the parameter level and loss level for a fixed operating point and different scaling factor values, on the loss level for fixed scaling factor values over a drive cycle, and on the efficiency level for a fixed scaling factor and three different operating points, all to satisfaction.
After validation, we framed a design optimization problem, minimizing the energy provided to the electric motor.}
We showcased our scaling laws with a sizing optimization study for a small set of topologies equipped with radial and axial flux motors, which showed that the all-wheel drive AFM topology is the most energy efficient, but also represents the highest material and production cost.
These results confirm the energy-efficient potential of AFMs, proving the need to considerably lower the cost of the technology, which could be achieved by leveraging learning rates and the economies of scale.

In future work, we aim at validating the scaling laws on the component level, for instance by comparing efficiency or loss maps \revVPPC{for the full operating envelope}, and investigating the impact of performance requirements on AFM size, efficiency and cost.
\revVPPC{Furthermore, this framework can readily be employed to evaluate more topologies and be extended to include more detailed material and production cost models.}

\addtolength{\textheight}{0cm}   


\section*{Acknowledgment}

We thank Dr.~I.~New for proofreading this paper.



\bibliographystyle{IEEEtran}
\bibliography{../../../Bibliography/main,../../../Bibliography/SML_papers} 

\end{document}